# Cumulative energy effect in the shock-discontinuity interaction under real-gas conditions

**A. Markhotok**

*Physics Department*
*Old Dominion University*
*Norfolk, VA 23529*

A recently proposed technique for improving ignition timing is investigated for real-gas effects. A correction to the model describing the cumulative energy effect in the presence of the shock-discontinuity interaction was done by re-considering the *RH*-relations and the shock refraction equations. Non-dissociating gas in thermal equilibrium state, with excited translational, rotational, vibrational, and electronic degrees of freedom is considered. When applied to $N_2$ gas at *T*=3000 *K*, 11% density increase and 17% temperature and 7% pressure jump decrease were found, compared to that in ideal gas. The derived relations are largely built on experimental data thus avoiding the complexity of careful accounting for all real gas effects and still offering satisfactory precision levels. The main result is that in real gases the effect in the interaction can be stronger than that in an ideal gas. The findings can be of interest in design of efficient combustors for hypersonic systems, detonation, space and craft design, shock-flame interaction, in astrophysics, and fusion research.

E-mail of corresponding author: amarhotk@phys.washington.edu

## I.     Introduction

The topic of shock wave assisted combustion, when a shock wave is employed to improve the ignition conditions in the gas, is of interest due to its applications for plasma-based ignition, flight control, and combustion enhancement. Based on this, techniques help enhance the performance of hypersonic vehicles powered by scramjets, efficient combustors for hypersonic propulsion systems, gas turbine and diesel engines, spark inhibition, for explosion limits in blended fuels and to make combustors more environmentally friendly and economical. In some of the applications, such as air breathing propulsion at supersonic speeds, the combustion performance can be a subject of strict requirements.  At about 1000 *K* and 0.5-2.0 atmospheres, the component mixing, ignition, and combustion should proceed in the micro- to millisecond time scale range. Such fuels as liquid hydrocarbons, for example, offer excellent characteristics associated with high energy density, storage stability, cooling capability, and compatibility with the existing fuel infrastructure. However, it does not satisfy the key characteristics in combustion timing, and thus is dependent on combustion enhancement techniques [1]. Such techniques, essentially based on inducing and promoting ignition, are working through shortening the ignition delays, and thus applicable to the situations where ignition would not otherwise occur.

The hot gas piloting, photo irradiation, plasma and laser assisted ignition, and heterogeneous and homogeneous catalysts are among the current and successfully working techniques [2-6].

Another technique capable of improving ignition timing employs the effects of interaction between a shock wave and plasma with a specifically decreasing density distribution [7]. A shock wave moving through such a medium can accelerate without a limit [8] (the cumulative energy effect) thus facilitating an expedited increase in local gas temperature to initiate the ignition. The ability to achieve very high gas temperatures almost instantly combined with precise locality could be a key to solve the timing problem in the ignition applications. The shock's ability of reaching extremely high speeds, up to sub-relativistic, could also be a subject of interest in the nuclear fusion experiments where very high compressions/temperatures are sought.

Shock waves interacting with a heated neutral or weakly ionized gas feature a number of interesting phenomena affecting its structure and dynamic parameters, such as speed and direction [9-12]. Significant shifts in the timing and the gaseous media state are observed, as the continuing evolution in the shock structure and post-shock flow can still be observable after the discharge is off or after the shock leaves the discharge [13]. Depending on the state of gas, geometry of the interaction, and the gas parameter distribution, the shock can accelerate [14] or decelerate until its full stop [15]. During the interaction, the front can gain/loose additional dimensions when changing from plane to curved and back [16-21]. The specific details of the gas state, such as electron kinetic, ionization/dissociation processes, radiation, atomic/molecular transitions, energy exchange processes in collisions, chemical processes, boundary effects, heat, charge (in plasma), or radiation convection [20] adds to the variety and complexity of the final state of the gas/plasma or the shock wave structure after the interaction. To dip a little bit more into the complexity, this research will consider a problem of real-gas effects on the interaction and their influence on the cumulative energy effect in the presence of gaseous discontinuity. In terms of applications, it is important to see whether real-gas properties can significantly alter the effects found in ideal gas [7] and how much it may limit or facilitate their use for improved ignition conditions.

A problem of a plane shock wave interacting with a heated gas/plasma medium with decreasing density profile will be considered. It will be shown that the real-gas properties will affect the main relations describing the interaction mostly through the Rankine–Hugoniot ($R–H$) relations that are used as a base for derivations. In the following paragraphs, first, a correction to the ideal-gas $RH$-relations will be determined, in general and then numerically. The choice of essentially contributing real-gas processes will be based on the specific application conditions brought up here. Then, the corrected $RH$-relations will be used together with the real-gas shock wave refraction equation that determines the relationship between the transmitted (refracted) and reflected off the interface energies. To get an idea of a possible size of the effect, the derived relations will be applied numerically to a specific problem involving cumulative energy effect in the presence of shock-interface interaction in diatomic nitrogen at temperature of 3000 $K$. The results are presented in the form of shock advancement profiles and its speed at a number of interaction times compared to those in ideal gas, and recommendations for its applications are discussed in the conclusion section.

## II. **Shock wave jump conditions in real gas**

In this paragraph, the attention will be given to determining the corrections to the shock wave jump conditions due to real-gas effects. The *R-H* relations determine gas parameter jump across the shock and, in its classical form, assume the ideal-gas conditions on both sides of the shock-front discontinuity. In their formal derivation, the shock wave is approximated with a singular, infinitely thin surface of infinitesimal rise time that mathematically represents a very narrow region across which sharp changes of final amplitude in the gas properties occur. Since dissipation is neglected in the derivation, any information on its mechanisms is not included in the equations and thus the specific entropy in the flow on both sides of the discontinuity must be constant. On the other hand, the compression in a shock wave is an irreversible process leading to an increase in entropy. To avoid discrepancy, the shock is allowed to have a microscopic structure and it is supposed that the entropy increase across the shock occurs within its layer of a finite width. The type of the dissipation process determines the width that is adjusted in such a way that the specific entropy increase across the shock corresponds exactly to the parameter jump across the shock satisfying the *R-H* relations. In case of two-body collisions typical for moderate gas conditions, the shock thickness is on the order of the molecules mean-free-path in the gas, however in some cases the shock layer can be up to tens of times wider.

In problems involving shock waves interacting with substantially heated media, the validity of the ideal gas approximations and applicability of the R-H relations must be questioned. The non-ideal gas effects are known to be quite strong, as for example in $O_2$ and $N_2$ where at very low pressures dissociation is significant. The density jump across a shock in those gases can, for example, exceed the maximum ideal diatomic value of 6 by several times reaching a value of 25 [21]. If an experiment is creating substantially non-perfect gas conditions and its results are used for modeling, the criterion of ideality has to be applied and tested. Thus accounting for non-perfect gas effects is very important and any data on particular gases or calculation procedures is much desired by researchers.

The research on this subject has already been done for various problems of gas dynamics, for example [22-27]. In this paragraph, the real-gas jump conditions will be determined for a range of temperatures common for gas discharges. Then the relations will be applied to a specific gas (diatomic nitrogen) and the results are presented in the form of graphs, compared to that predicted for an ideal gas.

In deciding what kind of mechanisms need to be accounted for when considering shock–plasma interaction, the presence or not of thermal equilibrium in the gas dynamic flow is the key factor. The energy transfer on an atomic level via ionization and dissociation, chemical reactions, electronic excitation, and radiation typically contribute in non-equilibrium state. The pure thermal mechanisms are more common for the flows that achieve at least local equilibrium [19]. The local equilibrium state is typical for at the last stages of decaying plasmas or pulsed laser energy deposition in quiescent air. The partial (by component) equilibrium can be present in plasmas generated by localized energy deposition, when charged and neutral particle ensembles depart from equilibrium but the bulk non-reacting gas is in equilibrium.

Thus, for the considered application conditions, in the derivations below we will assume that the system is in the thermodynamic (*TD*) equilibrium, and the pressure and temperature levels suppose no molecular dissociation in the gas. The time it takes for *TD* equilibrium to be established depends on rates at which transfer of energy between the degrees of freedom occurs. The times for translational and rotational degrees are very short, but for vibrational and dissociation degrees they can be several orders of magnitudes longer. The $CO_2$ gas is known, for example, for having a vibrational relaxation time relatively shorter than the dissociation relaxation time, so the vibrational equilibrium is established well before the dissociation comes into play. Thus the approach used here can be valid for the cases when the flow duration/interaction times are shorter than the dissociation time or for certain gases (for ex. $N_2$, $CO$) whose dissociation energies are high enough so this degree of freedom is not significantly excited.

In monatomic gases, the ideal-gas jump relations work until approximately 8000 *K* and at some lower temperatures in diatomic gases, when only translational and rotational degrees of freedom are excited. At higher temperatures, when the heat capacity of the gas increases due to excitation of vibrations, the temperature jump values $T_2/T_1$ begin to lag behind its ideal-gas equivalent. This lag experiences steep increases with the onset of dissociation, and then electronic excitation and ionization as the temperature goes up. The real-gas density ratio $\rho_2/\rho_1$ in a shock wave considerably overtakes the ideal-gas levels and this tendency increases with the shock intensity, while the pressure jump usually becomes slightly elevated.

In one of the existing models very accurately accounting for real-gas effects, its relations rely on the values of the shock speed that are measured experimentally [21]. Unfortunately it can't be used fully for the problem considered here because relations developed in the next paragraph will require analytical expressions as functions of the shock strength (Mach number). Thus, for the purpose of present work, the *RH*-relations must be re-derived in the required form.

Derivation of real-gas jump relations can be started with the classical expressions of the three fundamental laws of conservation (mass, momentum, and energy) that are still valid for real gas

$$\rho_1 u_1 = \rho_2 u_2 \tag{1}$$

$$p_1 + \rho_1 u_1^2 = p_2 + \rho_2 u_2^2 \tag{2}$$

$$H_1 + \frac{1}{2}u_1^2 = H_2 + \frac{1}{2}u_2^2 \tag{3}$$

Here subscripts *1* and *2* correspond to the flow in front and behind the shock correspondingly, *H* is the specific enthalpy, $u_2 = V_{sw} - v_2$, $u_1 = V_{sw} - v_1$ are the flow speeds across the shock in the reference frame moving with the shock wave, $v_1$ is the flow speed in front of the shock relative to laboratory reference frame, and $V_{sw}$ is the shock speed relative to the gas. Using the first two equations in the system, the energy conservation equation (3) can be brought to the form

$$H_2 - H_1 = \frac{u_1^2}{2}\left[1 - \left(\frac{\rho_1}{\rho_2}\right)^2\right] \qquad (4)$$

Since the heat capacities $C_p$ and $C_v$ are not constant in real gases but are functions of temperature and pressure, the enthalpy cannot be written in its classical ideal-gas form and the system (1-3) cannot be solved explicitly. Thus, the real-gas solution can be reduced to finding a way to express the enthalpy in terms of real-gas parameters and the shock wave strength.

The complexity of careful accounting for various real-gas processes in the left-hand part of the expression (4) can be avoided if experimental data obtained for specific gases is used. Such data on real-gas enthalpy for various gases is available in literature, for ex. [21]. The numerical data for the enthalpy can be fit with a polynomial function of temperature

$$H(T) = \sum_{n=0}^{\infty} \alpha_n T^n \qquad (5)$$

Because of the enthalpy difference in (4), only the linear ($\alpha_1$) and higher order coefficients $\alpha_n$ need to be determined. When applied to diatomic nitrogen, the fit (5) turns out to be quite linear. So if the higher order coefficients $\alpha_n$ are neglected, the enthalpy $\Delta H \approx \alpha_1(T_2 - T_1)$. Then the eq. (4) transforms into

$$\frac{T_2}{T_1} = 1 + \frac{u_1^2}{2*10^7 J\alpha_1 T_1}\left[1 - \left(\frac{\rho_1}{\rho_2}\right)^2\right] \qquad (6)$$

where "J" is the mechanical equivalent of heat, $u_1$ is in *cm/s*, and the enthalpy is in *Cal*. For the compressions achievable in shock waves, the ideal-gas state equation

$$\frac{p_2}{p_1} = \left(\frac{\rho_2}{\rho_1}\right)\left(\frac{T_2}{T_1}\right)\frac{\mu_1}{\mu_2} \qquad (7)$$

can be still valid [21]. Here $\mu_1$ and $\mu_2$ are the gas molar weights and $d = \mu_2/\mu_1$ is the dissociation. Using this equation together with the system (1-3) yields

$$\frac{P_2}{P_1} = 1 + \frac{\mu_1 u_1^2}{RT_1}\left[1 - \frac{\rho_1}{\rho_2}\right] \qquad (8)$$

and combining (6) , (7) and (8) we obtain

$$1 + \frac{\mu_1 u_1^2}{RT_1}\left[1 - \frac{\rho_1}{\rho_2}\right] = \frac{\rho_2}{\rho_1}\frac{\mu_1}{\mu_2}\left\{1 + \frac{u_1^2}{2*10^7 J\alpha_1 T_1}\left[1 - \left(\frac{\rho_1}{\rho_2}\right)^2\right]\right\} \qquad (9)$$

The speed $u_1$ in the eq. (9), which is equal to the shock speed relative to the gas (at $v_1 = 0$), can be expressed through the speed of sound $a_1$ and the shock Mach number $M_1$, $u_1 = M_1 a_1$. Thus the equation (9) can be solved for the density ratio versus $M_1$. In determining the speed of sound, the real gas processes must be accounted for as well. This will be done in the next paragraph, in the form of correction factor $\sigma$ used with the ideal-gas relation as $a_1^2 = \sigma(\gamma_i RT/\mu)$. For diatomic nitrogen, the dissociation levels remain insignificant even at *T*=5000 *K* and thus the ratio $d = \mu_2/\mu_1$ in eq. (9) can be neglected. Then, with known solution for the density ratio obtained from the eq. (9), the full set of jump relations (eqs. 7, 8, 9) can be obtained. This system of equations will be used to replace the ideal gas Rankine-Hugoniot relations.

Since the experimental data on enthalpy for a particular gas is available in a definite range of temperatures, the use of the solution (7-9) is limited by this range. In case the data is outside of a desired temperature interval, similar results for diatomic gases can be obtained with the following alternative solution. The enthalpy of a real gas consisting of diatomic molecules in equilibrium state at a temperature $T$ can be written as

$$H_T = \frac{3}{2}RT + \frac{2}{2}RT + \frac{\bar{N}h\nu}{(e^{h\nu/kT}-1)} + RT + \frac{h\nu}{2} \qquad (10)$$

where the first three terms correspond to the excited translational, rotational and vibrational degrees of freedom respectively, the last one is the zero point energy term, and no dissociation or electronic excitation is taken into account [21]. Here $\bar{N} = N/\mu$ is the number of molecules per unit mass of gas, $R$ is the universal gas constant per unit of mass of gas, and ν is in cycles/second. If no dissociation ($N_2 = N_1$),

$$H_2 - H_1 = \frac{7}{2}R(T_2 - T_1) + Nh\nu\left[\frac{1}{(e^{h\nu/kT_2}-1)} - \frac{1}{(e^{h\nu/kT_1}-1)}\right] \tag{11}$$

The vibrational energy term represents the summation over $N$ quantized molecular oscillators having permissible energy levels corresponding to frequencies $\nu_n = n\nu$, with the fraction of molecules in any level being proportional to $\exp(-\frac{h\nu}{kT})$. This applies to the case of a diatomic molecule in its electronic ground state vibrating with a characteristic frequency ν measured in cycles per second. The zero point energy term $h\nu/2$ does not enter the relation (11) since it does not depend on temperature. For nitrogen, at the temperature of $T = 3000\ K$ and the characteristic frequency $\nu = 0.826 \cdot 10^{13}\ s^{-1}$ the ratio $h\nu/kT \ll 1$ so the exponential term can be approximated [28] with the first three terms in the expansion $e^{\frac{h\nu}{kT}} \cong 1 + \frac{h\nu}{kT} + \frac{1}{2}\left(\frac{h\nu}{kT}\right)^2 + \cdots$. Then, using the same approach as in the previous derivation, the eq. (4) and (10) can be used to obtain

$$u_1^2\left[1 - \left(\frac{\rho_1}{\rho_2}\right)^2\right] = \frac{7}{2}RT_1(\frac{T_2}{T_1} - 1) + RT_1\left[\frac{\frac{T_2}{T_1}}{1 + \frac{1}{2T_1}\left(\frac{h\nu}{k\frac{T_2}{T_1}}\right)} - \frac{1}{1 + \frac{1}{2}\left(\frac{h\nu}{kT_1}\right)}\right] \tag{12}$$

If solved together with the relation (6), this yields a dimensionless equation that can be solved for the temperature ratio $T_{21} = T_2/T_1$

$$\frac{T_{12}d}{\sqrt{b_5 - b_1 T_{12} - \frac{b_2 T_{12}^2}{a_3 + T_{12}}}} = 1 + b_6\left[1 - \sqrt{b_5 - b_1 T_{12} - \frac{b_2 T_{12}^2}{b_3 + T_{12}}}\right] \tag{13}$$

The solution to the equation (13) is

$$\frac{T_{12}d}{s} = 1 + b_6[1 - s] \tag{14}$$

where $s = \sqrt{b_5 - b_1 T_{12} - \frac{b_2 T_{12}^2}{b_3 + T_{12}}}$, $d$ is dissociation, and the coefficients $b_1 = \frac{7/2}{\gamma M_1^2 \sigma^2}$, $b_2 = \frac{1}{\gamma M_1^2 \sigma^2}$, $b_3 = \frac{h\nu}{2kT_1}$, $b_4 = \frac{17/2}{1+b_3}$, $b_5 = 1 + \frac{9/2 + \left(1 + \frac{h\nu}{2kT_1}\right)^{-1}}{\gamma M_1^2 \sigma^2}$, $b_6 = \gamma M_1^2 \sigma^2$ are the functions of the complex dimensionless parameters $\gamma M_1^2 \sigma^2$ and $h\nu/2kT_1$. Here $\gamma$ is the real-gas adiabatic coefficient and $\sigma$ is the correction factor to the speed of sound (discussed in next paragraph). The solution for $T_{21}$, together with (6-8) represent the alternative real-gas system of R-H relations that can be used to determine the temperature, density and pressure jumps across the shock.

### III.     Speed of sound in real gases

To estimate real-gas effects on the speed of sound, the approach [29] using a correction to its ideal gas value can be utilized. The real-gas equation will keep the Laplace equation form $a^2 = \gamma_i RT/\mu$ but corrected with three factors

$$a^2 = (\gamma_i RT/\mu)(1 + K_c)(1 + K_v)(1 + K_r) \qquad (15)$$

Here the $i$ - subscript refers to the ideal gas where only translational and rotational degrees of freedom are excited, the compression factor $z = P/\rho RT$ is independent of pressure, there are no losses during the sound wave propagation and the parameter $\sigma = (1 + K_c)(1 + K_v)(1 + K_r)$. The coefficient $K_c$ is the specific heat correction accounting for the temperature. $K_v$ is the virial correction taking into account intermolecular interactions and is the function of temperature and pressure. And the relaxation correction $K_r$ accounts for relaxation processes leading to acoustical dispersion and is a function of temperature, pressure and frequency. All the three corrections are independent of each other and thus can be calculated separately.

### III A.  Specific Heat Correction

One of the ways to estimate the specific heat correction $K_c$ can be the use of a fit function to experimental data that is available, for example, in reference [29]

$$K_c = \frac{\gamma_c}{\gamma_i} - 1 = \frac{1}{\gamma_i}(1 + a_{c0} + a_{c1}T + a_{c2}T^2 + a_{c3}T^3 + a_{c-}T^{-1}) - 1 \qquad (16)$$

The applicability of the equation (16) is dependent on the range of temperatures within which the data is available. For example, for diatomic nitrogen the temperature range in the above reference is relatively narrow, between 80 and 775 $K$. If higher temperatures are of interest, an alternative procedure using relations from [29] can be used. Each type of the various excited modes in the gas are considered there separately assuming that the heat capacity $C_{vc}$ is additive. The Decoupled Ridged-Rotor Harmonic oscillator (*DRH*) model that can be used in this case is based on the assumption that there is no Coriolis interaction between the modes, no modes of motion associated with molecular distortions are excited and no low-$T$ quantum effects. The model precision is on the order 0.5% for temperatures up to 1500 $K$ and the number increases with further temperature increase. The square well potential model can also be applied to the problem though its precision was shown to be much less than for the *DRH* model, with up to +/- 20% error.

The following mode contributions to the specific heat will be considered, in the typical order of their excitation as the gas temperature increases. The translational specific heat $C_{v0} = \frac{3}{2}R$, if no other modes are present, results in the contribution to the correction coefficient

$$K_c = \frac{\gamma_c}{\gamma_i} - 1, \quad \gamma_c = 1 + \frac{R}{C_{vc}} \qquad (17)$$

The rotational specific heat (for linear molecules)

$$\frac{C_{vc,rot}}{R} = 1 + \frac{1}{45}\left(\frac{\theta_{rot}}{T}\right)^2 + \cdots \qquad (18)$$

and for nonlinear molecules

$$\frac{C_{vc,rot}}{R} = \frac{3}{2} + \frac{1}{45}\left(\frac{\theta_{rot}}{T}\right)^2 + \cdots \qquad (19)$$

where the values of characteristic temperature $\theta_{rot}$ in the quantum correction (second) term are available [29]. For the air gas components such as $N_2$, $NO$, and $O_2$, for example, the characteristic temperature values are very low, 2.81 $K$, 2.45 $K$, and 2.08 $K$ respectively, making the correction, even at room temperature, negligible.

The vibrational specific heat for a diatomic molecule can be estimated using the harmonic oscillator mode

$$\frac{C_{vc,vib}}{R} = \left(\frac{\theta_{vib}}{T}\right)^2 \frac{exp\left(-\frac{\theta_{vib}}{T}\right)}{\left[1-exp\left(-\frac{\theta_{vib}}{T}\right)\right]^2} \qquad (20)$$

where $\theta_{vib}$ is the characteristic temperature [29].

For the conditions developed in shocks, the typical contribution of electronic excitation to the specific heat

$$\frac{C_{vc,el}}{R} = g_0 g_1 \left(\frac{\theta_{el}}{T}\right)^2 \frac{exp\left(-\frac{\theta_{el}}{T}\right)}{g_0 + g_1 exp\left(-\frac{\theta_{el}}{T}\right)} \qquad (21)$$

is usually very insignificant due to it's high (relatively to the previous modes) excitation energies, except in some specific gases. For example, in $N_2$ molecule, its singlet ground $X_1 \Sigma_g{}^+$ and first excited a $^1\Pi_g$ states are separated by 69290 cm$^{-1}$ that is corresponding to $\theta_{el} = 69{,}290$ $K$. With both degeneracies $g_0 = g_1 = 1$, the correction, at room temperature, yields just the value $2.4 \cdot 10^{-143}$ that can be safely neglected. However, in the $NO$ molecule, for example, the ground level is doubly degenerate: the spin-orbit coupling splits the ground level ($^2\Pi_{3/2}$ and $^2\Pi_{1/2}$ states) with the separation of only 174.2 $K$ that atypically strongly contributes to the specific heat [29].

Using the specific heat additivity, the contribution of all modes adds up

$$\frac{R}{C_{vc}} = \left(\frac{C_{vi}}{R} + \frac{C_{vc,rot}}{R} + \frac{C_{vc,vib}}{R} + \frac{C_{vc,el}}{R}\right)^{-1} \qquad (22)$$

that can be used in (17) to finally obtain the correction $K_c$.

### III B. Virial Correction

The virial correction factor $K_v$ accounts for intermolecular interactions. For compressions typical in shocks its value is rather negligible, however in some media with specific molecular structure and characterized by low temperatures and high pressures, it should be considered. The virial correction can be determined using the expansion of the equation of state with pressure $P$ used as the expansion variable

$$\frac{PV}{RT} = 1 + B_p P + C_p P^2 + \cdots \qquad (23)$$

Here $V$ is the molar volume and $B_p = B/RT$ and $C_p = (C-B^2)/(RT)^2$ are the second and third pressure virial coefficients related to the corresponding volumetric coefficients $B$ and $C$. Then the virial correction, if limited with the first two terms,

$$K_v = FP + GP^2$$

where $F = \frac{K}{RT}$, $G = (L - BK)/(RT)^2$ and the acoustical virial coefficients $B$, $K$ and $L$ will be determined below as follows. The coefficient $B$ can be obtained by fitting experimental data, as described in reference [29].

$$B = a_v - b_v exp\left(\frac{c_v}{T}\right) \tag{24}$$

The coefficient $K$ can be expressed via the following fit

$$K = a_{v0} + \left(a_{v1} + \frac{a_{v2}}{T} + \frac{a_{v3}}{T^2}\right) exp\left(\frac{c_v}{T}\right) \tag{25}$$

and the numerical coefficients in (25) are related to those in (24) as

$$a_{v0} = 2a_v, \ a_{v1} = -2b_v, \ a_{v2} = \frac{2(\gamma_0-1)c_v b_v}{\gamma_0}, \ a_{v3} = \frac{(\gamma_0-1)^2(c_v)^2 b_v}{\gamma_0} \ .$$

The coefficient $L$ can be determined in a similar way, using a fit to the data for the third virial coefficient. Due to bulkiness of the final expressions, they are not presented here but instead can be found in [29].

Molecular association that is also related to the virial coefficients is impacted through the molar mass, specific heat $C_{vc}$, and the compressibility factor [29]. Molecular collisions can produce stably bound long-lived or metastably bound short-lived molecular formations that should be considered depending on whether the system is in $TD$ equilibrium or not. For linear molecules, the association of two monomers result in the loss of 3 translational and 3 rotational degrees of freedom that are replaced by 6 dimer-specific vibrational, "weak" modes. If the mode's characteristic temperatures are relatively low, they contribute $6R$ to the specific heat, so the total specific heat $C_{V,v} = x_1 3R + x_2(3R + 6R) = (3 + 6x_2)R$ and the specific heat ratio $\gamma_v = (4 + 6x_2)/(3 + 6x_2)$. Here $x_1$ and $x_2$ are mole fractions for monomers and dimers correspondingly, and $x_1 + x_2 = 1$.

### III C.  Relaxation Correction $K_r$

Sound wave energy dissipates via collisions during which a portion of energy is transferred to the translational degrees of freedom followed with its transfer to the inner molecular degrees of freedom. During the relaxation period gas parameters change until a new $TD$ equilibrium establishes. Because it may take up to billions of collisions to activate each of the transitions and, accounting for finite de-excitation times, the time delay between excitation and response causes the sound wave dispersion followed by change in the speed of sound and absorption. If impurities are present in the gas, the dispersion effect can be even stronger.

The sound speed $a$ and the absorption $\mu_1$ can be expressed in terms of relaxation strength $\varepsilon$, frequency $\omega$, and relaxation times, the dispersion time $\tau_d$ and absorption time $\tau_a$ [29], as

$$a^2 = a_{cv}^2 \left[1 + \frac{\varepsilon}{1-\varepsilon} \frac{(\omega\tau_d)^2}{1+(\omega\tau_d)^2}\right] \tag{26}$$

$$\mu_1 = \pi \frac{\varepsilon}{\sqrt{1-\varepsilon}} \frac{\omega\tau_d}{1+(\omega\tau_d)^2} \tag{27}$$

Here the subscript $i$ will be referred to an ideal gas and $c$ - to a lossless real gas that accounts for two previously discussed corrections, $a_{cv}^2 = a_i^2(1 + K_c)(1 + K_v)$, and times $\tau_d$ and $\tau_a$ are related as $\tau_d = \tau_a\sqrt{1-\varepsilon}$. Then the relaxation correction

$$K_r = \frac{\varepsilon}{1-\varepsilon} \frac{(\omega\tau_d)^2}{1+(\omega\tau_d)^2} \qquad (28)$$

The relaxation strength $\varepsilon$ for a specific $i$-th reaction can be obtained via the corresponding specific heat capacity $C_i$ and using its additivity property

$$\varepsilon = \frac{R\sum_i C_i}{(C_p^0 - \sum_i C_i)(C_p^0 - R)} \qquad (29)$$

If an $i$-th vibrational degree of freedom is excited, its isothermal heat capacity can be determined as

$$C_i = q_i R \left(\frac{\theta_{vib}^{(i)}}{T}\right)^2 \frac{exp\left(-\frac{\theta_{vib}^{(i)}}{T}\right)}{\left[1 - exp\left(-\frac{\theta_{vib}^{(i)}}{T}\right)\right]^2} \qquad (30)$$

where $R$ is the universal gas constant, $g_i$ is the vibrational energy level degeneracy, and the vibrational temperature values $\theta_{vib}$ are found experimentally and listed for various gases in [29].

The relaxation time $\tau_{VT}$ can be calculated using the following fit to experimental data

$$\log(\tau_{VT}P) = a_r(1) + a_r(2)T^{-\frac{1}{2}} + a_r(3)T^{-1} \qquad (31)$$

where $T$ is the temperature in $K$, $P$ is pressure in $atm$, time is in $\mu s$, and the numerical coefficients $a_r(i)$ are given in [29]. Then the dispersion time can be found using the values of time $\tau_{VT}$, specific reaction heat capacity $C_i$ and that of an ideal gas $C_V^0$

$$\tau_d = \tau_{VT}\left(1 - \frac{C_i}{C_V^0}\right) \qquad (32)$$

The range of sound wave frequencies $\omega$ is specific to the way it is created. However, the correction $K_r$ in the equation (28) can be readily determined for two limiting cases. When the complex $\omega\tau_d \ll 1$, the correction simplifies to

$$K_r = \frac{\varepsilon}{1-\varepsilon}(\omega\tau_d)^2 \qquad (33)$$

and this condition corresponds to the frequencies

$$\omega_k \ll \left[\tau_{VT}\left(1 - \frac{C_i}{C_V^0}\right)\right]^{-1} \qquad (34)$$

If $\omega\tau_d \gg 1$, the correction $K_r = \frac{\varepsilon}{1-\varepsilon}$ is the function of the relaxation strength only, and applies to the frequencies $\omega \gg \omega_k$.

### III D. Correction to the speed of sound in diatomic nitrogen

As a numerical example, the real-gas correction to the speed of sound will be determined here for diatomic nitrogen. This specific gas was chosen due to its dominant contribution in the air and the gas temperature taken as $T_2$=3000 $K$ is common for discharges. For these specific parameters, the approach using eq. (16) cannot be used because the numerical coefficients $a_{ci}$ in eq. (16) are available only in the range of temperatures between 80 $K$ and 775 $K$. Then the alternative procedure with equations (17-32) will be utilized. In the calculations below, all experimental data including the values of characteristic temperatures will be taken from reference [29] unless otherwise noted.

To determine the value of specific heat correction $K_c$, the contribution of the following degrees of freedom in the gas will be taken into account. In the nitrogen gas, the translational degree of freedom term $C_{vi}/R = 3/2$. The value of rotational specific temperature in the quantum correction (second) term $\theta_{rot} = 2.08\ K$ is very low. This makes the correction to the rotational specific heat in eq. (18), $\frac{C_{vc,rot}}{R} = 1 + 2.0\cdot10^{-8}$ negligible. However the characteristic vibrational temperature in this gas, $\theta_{vib} = 3352.0\ K$ is comparable to the gas temperature $T_2$ and the correction to the specific heat term in eq. (20), $\frac{C_{vo,vib}}{R} = 0.902$ is significant. For electronic transitions, the characteristic temperature $\theta_e = 99{,}692\ K$ is relatively high and in the ground state ($g_0 = g_1 = 1$) the contribution $C_{vc,el}/R = 4.14\cdot10^{-7}$ can be neglected. As temperature increases, its value is still low reaching $4.5538\cdot10^{-3}$ at $T = 10{,}000\ K$.

The contribution of all modes into the total specific heat can be determined using the additive property of heat capacities, $\frac{C_{vc,Total}}{R} = 3.402$, so $K_c = 0.0758$ and $\gamma_c = 1 + \frac{R}{C_{vc,T}} = 1.293$. Applying the procedure within a range of temperatures 300 -10,000 $K$, the correction $K_c$ vs temperature curves can be obtained. The results presented in Fig. 1 can be used as an extension into higher temperatures of similar results obtained in reference [29].

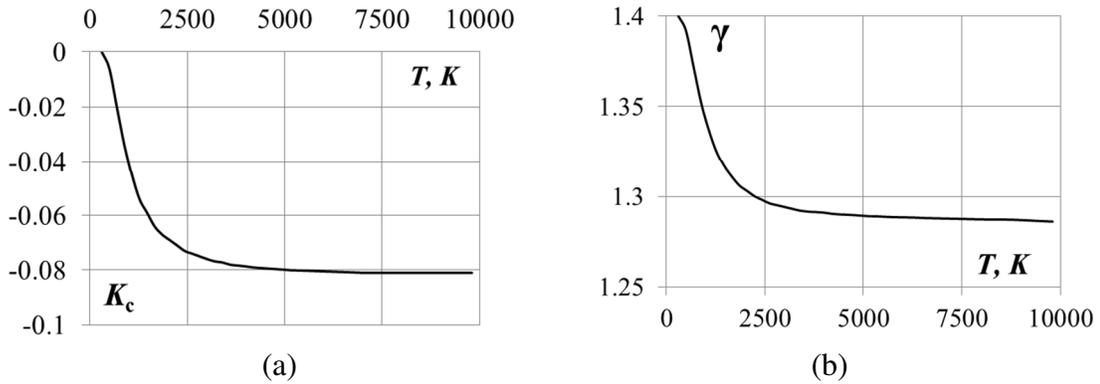

*Fig. 1. Specific heat correction $K_c$ (graph a) and corresponding specific heat ratio $\gamma_c$ (graph b) vs gas temperature, for $N_2$ gas.*

For the virial correction, a very small value $K_v = 0.374\cdot10^{-8}$ can be obtained using the equations (23-25), and thus, in considered circumstances, its contribution to the sound speed is insignificant.

The value of relaxation correction $K_r$ at the specified temperature can be determined using equations (26-32). In diatomic nitrogen, when only one vibrational degree of freedom is excited, its isothermal heat capacity $C_i = 7.5016\ J/(mol K)$ can be determined from eq. (30) at $\theta_{vib} = 3352.0\ K$. Then the relaxation intensity $\varepsilon = 0.04558$ is obtained from the eq. (29). Using numerical coefficients $a_1 = -3.6481$, $a_2 = 71.6300$ and $a_3 = 0$ in the equation (31), the time $\tau_{VT} = 20.827\ \mu ks$. Then, from the values of $\tau_{VT}$, specific heat capacity $C_i$, and that of an ideal gas $C_{vi}$ in the eqn. (32) the "relaxation" time $\tau_d = 16.56\ \mu ks$.

The sound wave frequency $\omega$ that is entering the expression (28) can be determined from experimental data. Particularly, when a shock wave interacts with a plasma volume created in an optical discharge, the frequency can be estimated from pressure rise time in the gas explosion. In one of such experiments [30], the typical time for the pressure to increase to maximum was around $t_0 = 50$ μks corresponding to the frequency $\omega = 12.56 \cdot 10^4$ *rad/s*. Then the correction $K_r = 0.0388$, and thus, with all the three corrections $K_c$, $K_v$, and $K_r$ known, the speed of sound correction can be determined as $\sigma = 0.9472$.

## IV. Shock wave jump relations in real gas

With the value of the speed of sound correction determined, the real gas jump relations can be computed using the system of equations (6-9). The curves in Fig. 2 are the jump relations versus Mach number, in non-dissociating ($d = 1$) real gas ($N_2$) obtained at the temperature $T=3000$ *K*, presented comparatively to the ideal gas case. The results represent real-gas replacement (solid lines) of the jump relations in ideal gases (dashed line). On the graphs, the green line corresponds to the speed of sound corrected results (σ = 0.9472). The black solid line corresponding to uncorrected speed of sound results (σ = 1) facilitates the size of the effect.

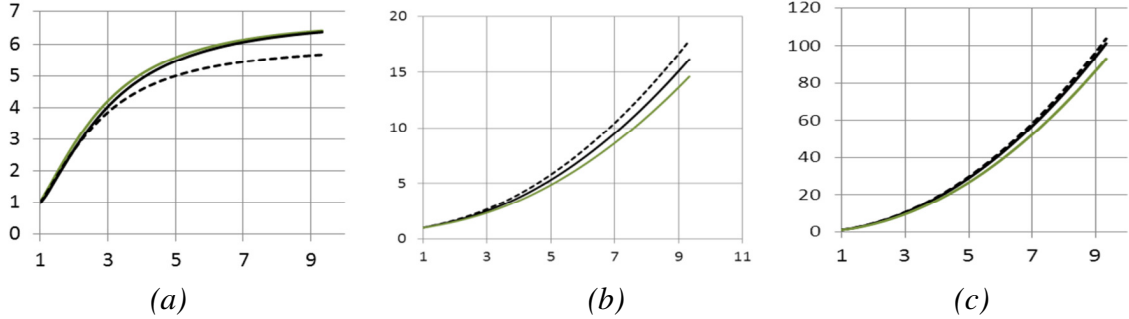

*(a)*                          *(b)*                          *(c)*

*Fig. 2. Shock wave jump conditions versus Mach number in non-dissociating real gas ($N_2$) at T=3000K, with the speed of sound corrected (σ = 0.9472, green curve) and uncorrected (σ = 1, black solid), comparatively to that in ideal gas (dashed line). (a) Density, $\rho_{21}$, (b) Temperature, $T_{21}$, and (c) Pressure, $P_{21}$.*

The results show noticeable departure of the jump relations from its ideal gas values. For this particular gas and temperature, around 11% increase in the density jump, 17% decrease in the temperature, and 10% decrease in pressure relative to the ideal gas values are found.

## V. Cumulative energy effect in the presence of a discontinuity under real-gas conditions: model and approximations

The cumulative energy effect is taking place when a shock wave propagates through a gas with decreasing to zero density distributed with a power law $\rho = \partial x^N$. Here $x$ is the distance to the zero density plane and *N* is the peripheral index of the polytropic curve [15, 31]. In the gas-dynamic approximation, a plane shock wave moving toward the zero density plane accelerates very quickly and is capable of reaching sub-relativistic speeds [31] The possibility of accumulation virtually infinite energy and unlimited temperature rise in the gas behind the shock is due to transfer of a finite amount of heat to a decreasing to zero quantity of matter. In reality though, as the shock reaches the very last gas layers, the macro approach becomes invalid. The supposedly infinite gas temperature at the zero density point will actually be limited by available and enough quick dissipative processes, usually radiation. This effect was first used to successfully explain radiation from novae, when a shock wave generated by supersonic movements from the internal disturbances inside a star emerges at the star surface [32]. It also explained the origin of cosmic rays, when during the explosion of the supernovae some of the material is ejected from the surface. The particles behind the shock wave accelerated to relativistic speeds become capable of overcoming the gravitation and escape from the star [32].

The specific density distribution can be found, for ex., on stellar surfaces in the form of heated plane or spherical layers in *TD* equilibrium in the presence of radiative transfer and radiant heat exchange. It is formed as the result of the combined action of gravity, thermal pressure, and radiant heat conduction atmospheres [33]. On the laboratory scale, the systems are formed based on the same combination of processes common for stellar atmospheres. Among the examples are: the gas of a radiative spherical cloud under simultaneous mechanical equilibrium and radiative transfer [32-38]; a gas layer with a finite mass and constant initial gas distribution suddenly expanding into vacuum [34,35]; sudden expansion of a spherical gas cloud into vacuum [33] originating from a strong explosion on a solid surface [34]; vaporization of the anode needle of a pulsed x-ray tube caused by strong electron impacts [36]; explosion of wires by electric currents in vacuum systems; spark discharge in air in the early stages [35]; the density distribution near the edge of a cooling wave [34]; motion of gas under the action of an impulsive load [36-38], a problem of bubbles collapsing in a liquid [34], and the spherical shock wave implosion [34,38,39].

A model of a shock wave interacting with plasma featuring a similar density distribution has already been developed for ideal gas [7]. In the present work, a model accounting for real-gas effects will be considered. As in the above reference, it will be assumed that a plane shock wave enters a spherically shaped plasma cloud from left to right, with velocity $V_1$ along the *x*-direction. The cloud is created in a discharge (high voltage, optical, etc.) and enough time is allowed for the gas to achieve TD equilibrium state. The gas density decreases to zero in the direction of the shock propagation and there is no density change in the transverse direction. The problem is considered in the vertical plane of symmetry that is normal to the incident shock plane. The density reaches its zero value at a plane *a* meters distant from the leftmost point at the interface. The discontinuous interface between the plasma cloud (medium 2) and the surrounding gas (medium 1) is assumed to be of spherical or cylindrical shape with the radius *R*. The temperature of cold surrounding gas $T_1$ in the first medium is assumed to be distributed homogeneously. $T_2$ is the hot plasma temperature right behind the interface that is changing with the distance starting at the leftmost point on the interface.

After encountering the cloud's interface, the shock will be partially reflected and the rest will propagate further through the cloud. Upon crossing the interface, the absolute value of the shock speed changes instantly from $V_1$ to $V_2$ and the total vector of the velocity rotates by the refraction angle $\gamma$. As a result, the front of the shock propagating through the plasma cloud becomes distorted in the form of increasing stretching along its motion direction.

The specific "progressive wave" ideal-gas solution [7] for the coordinates of a point (*i*) on the refracted front ($Xi, Y_i$) versus the incident's one ($x_i, y_i$) at a moment of time *t*

$$X_i = a_i - (G_i/b)(t_{0i} - t)^b, \quad Y_i = y_i - \overline{V} \sin\gamma \, [nR - x_i] \quad (35)$$

and the $x$-component of the shock velocity in the heated cloud

$$V_{2x} = V_1 \overline{V} \cos\gamma \left(\frac{t_{0i} - t}{t_{0i} - t_i}\right)^{b-1} \quad (36)$$

where the dimensionless time $n = t/\tau$, the characteristic time $\tau = R/V_1$ and the time

$$t_{0i} = \frac{b(a - R(1 - \cos\alpha))}{V_1 \overline{V} \cos\gamma} + n(1 - \cos\alpha) \quad (37)$$

The factor

$$G_i = \frac{V_1 \overline{V} \cos\gamma}{(t_{0i} - t_i)^{b-1}} \quad (38)$$

$\alpha$ is the incident angle (between normal to the interface and the $x$-direction), $\overline{V} = V_2/V_1$ is the dimensionless shock speed, $\overline{V} = \sqrt{\overline{T}\left(\overline{M}\right)^2 \cos^2\alpha + \sin^2\alpha}$, $\overline{T} = T_2/T_1$, $\overline{M} = M_{2n}/M_{1n}$ is the ratio of normal component of the Mach number, $k$ is adiabatic coefficient, $t_i = \tau(1 - \cos\alpha)$ is the time delay due to the interface curvature, the refraction angle $\gamma = \alpha - tan^{-1}\left(\sqrt{\overline{T}}M \tan\alpha\right)$, and coefficient $b = 0.59$ is borrowed from paper [15].

In the relations above, most of the parameters are determined by the problem geometry and arrangements. The ratio of normal components of Mach number $M_{2n}/M_{1n}$ accounting for the energy losses due to shock wave reflected off the interface [17], is the only parameter that is dependent on the gas state and thus is the subject to correction. The effects of gas non-ideality on the interface reflectivity has already been studied in [40], where the real-gas shock refraction equations were derived and used to determine a correction to the Mach number ratio. It was considered for a non-dissociating gas on both sides of discontinuity with excited translational, rotational, vibrational and electronic degrees of freedom. Thus the solution to the problem stated in this paragraph is reduced to determining a correction to the Mach number ratio and using it in the equations (35-38). Since the corrections are dependent on the sort of gas and its state parameters, it will be done in the form of illustration for a particular gas case.

## VI.  Numerical example for diatomic nitrogen at 3000 K

In this paragraph, a numerical example will be considered for a problem of a plain shock wave interacting with a thermal inhomogeneity under real-gas conditions. The intention is to estimate a possible size of real-gas effect on the energy cumulation in the interaction. The density distribution in the heated region and other parameters and approximations will be assumed the same as in the previous paragraph. Then the relations (35-38) can be used to keep track of the front development as it accelerates through the heated region. Since in the applications the ignition must be achieved in a timely manner and with precise locality, we will concentrate here on the shock speed in the heated area and the dynamics in shock front distortion.

Non-dissociating diatomic nitrogen will be considered as a gas medium on both sides of the interface, with the hot gas temperature $T_2 = 3000$ K, the temperature step $T_2/T_1 = 10$, and the incident shock Mach number $M_1 = 1.9$. The radius of the cylindrical interface is taken as $R = 0.3$ cm, and the specific heat ratio $k = 5/3$. For these parameters, all the relevant data on the gas and results for the Mach number ratio $M_{2n}/M_{1n}$ can be directly borrowed from ref. [40].

In accordance to the findings in that paper, at $T_2=3000$ K, the shift in the transition point determining the character of the reflected wave in real gas is much less than the heating intensity factor $T_2/T_1=10$. Consequently for a shock incident on an interface from-cold-to-hot medium, the reflected wave will be still a rarefaction wave (Fig. 3, a) and the following shock refraction equation must be chosen

$$\sigma \sqrt{\frac{\gamma_2}{\gamma_1}\frac{\mu_1}{\mu_2}\frac{T_2}{T_1}} M_2 \left(1 - \frac{\rho_1}{\rho_2}\right) = \frac{2\sqrt{\Gamma}}{\Gamma+1}\left(M_1 - \frac{1}{M_1}\right) - \frac{2\sqrt{\Gamma}}{(\Gamma+1)(\Gamma-1)}\left\{\left(2\Gamma M_1^2 - (\Gamma-1)\right)\left(\frac{2}{M_1^2}-(\Gamma-1)\right)\right\}^{1/2} \times$$
$$\left\{1 - \left[\frac{(\Gamma+1)(P_2/P_1)}{2\Gamma M_1^2 + (\Gamma-1)}\right]^{\frac{\Gamma-1}{2\Gamma}}\right\} \qquad (39)$$

In the equation, the data for real-gas corrected pressure $P_2/P_1$ and density $\rho_2/\rho_1$ jumps across the refracted wave (see Fig.2) and a correction to the speed of sound can be calculated using the procedures described above and in ref. [40]. With all those parameters in place, the eq. (39) then can be solved together with the system (35-38) to obtain the data on the shock profiles and the $x$-component of the velocity.

In the Fig.3b, numerical results for normal components of Mach number $M_{2n}$ for the refracted wave vs the incident one, $M_{1n}$, (solid line) compared to that in ideal gas (dashed line), were obtained using the equation (39). The elevated value of the Mach number compared to that in ideal gas is due to a decreased interface reflectivity that is possible in some real gases [40]. This is confirmed in the Fig.3a, where the levels of pressure jump across reflected waves (for both, shock and rarefaction waves) tend to be closer to the unity compared to that in ideal gas.

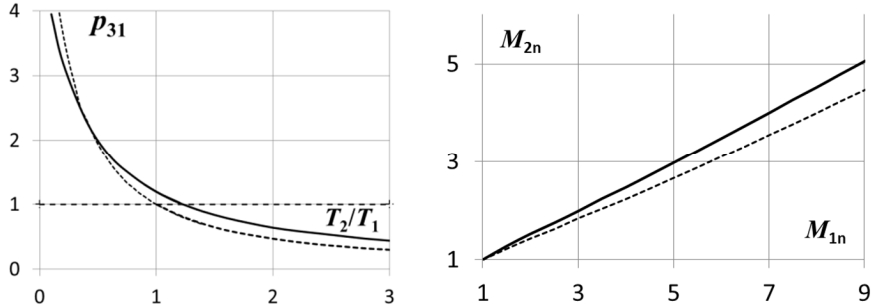

Fig. 3. (a) Pressure jump across reflected wave $p_{31}$ versus the temperature step $T_2/T_1$. (b) The transmitted normal component of shock Mach number $M_{2n}$ vs the incident one, $M_{1n}$, compared to that in an ideal gas. The gas is diatomic nitrogen at $T_2 = 3000$ K, $T_2/T_1 = 10$. Solid lines are for real and dashed ones – for ideal gas.

In the Fig. 4, the shock front profiles (graph $a$) and the $x$-component of the shock velocity $V_{2x}$ vs coordinate $y$ (graph $b$) are presented for real gas conditions (solid lines) compared to that in ideal gas (dashed lines).

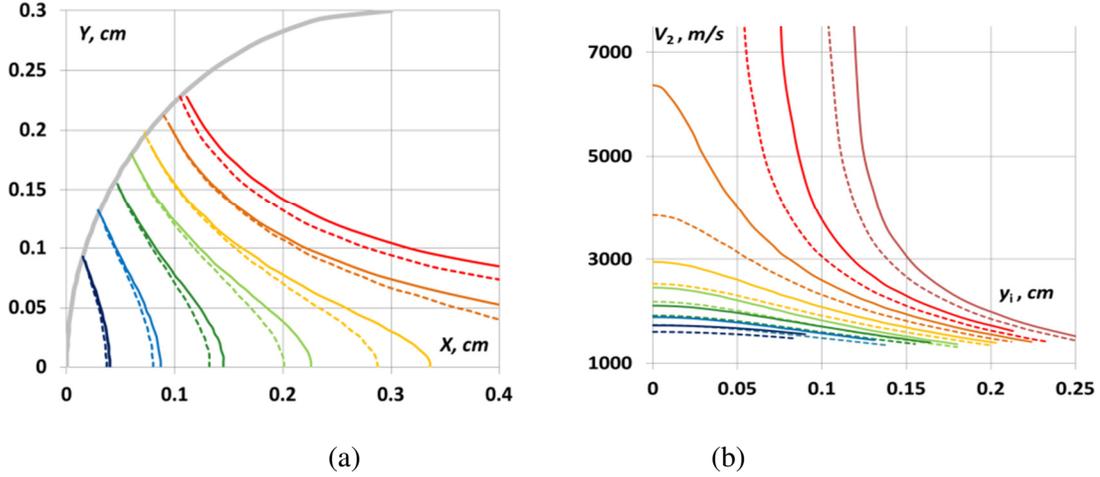

(a)　　　　　　　　　(b)

Fig. 4. Shock front profiles **(a)** and the *x*-component of the shock velocity $V_{2x}$ vs coordinate *y*, **(b)**, in diatomic Nitrogen at 3000 *K*, at times *n* between 0.05 and 0.35 trough equal intervals. The shock moves from left to right. Solid lines are for real- and dashed lines are for ideal-gas conditions. Each color corresponds to the same time. In graph (b), the time sequence is from lower to upper curves.

To have the cumulative energy effect prevailing over that associated with the shock refraction, the distance to zero density plane $a = 1.3R$ was taken comparable to the geometrical scale of the problem and long enough interaction times were allowed. To see the dynamics in the shock front distortion, the shock profiles were computed at a number of propagation times. For both graphs, the dimensionless times *n* corresponding to each curve are in the range between 0.05 and 0.35 through equal intervals and each color corresponds to the same time. In graph (*a*), the shock moves from left to right and the front portion that is still propagating in cold medium keeps its original shape (not shown). Due to symmetry, only the upper half of the graph is retained. In graph (*b*), the time sequence is from lower to upper curves.

As the graphs show, non-ideal properties of the gas considerably affect the interaction resulting in even quicker acceleration of the refracted shock and its front distortion. The data in graph (*b*) shows the shock speed values at the specified times. The physical reason explaining a stronger acceleration in real gas is associated with the decreased reflectivity off the interface. This results in a stronger transmitted shock that is additionally amplified when the shock is moving through the progressively rarefied gas layers.

## VII. **Summary and discussion**

A recently proposed technique for improved ignition timing [7] based on the use of the energy cumulation effect has been explored for real-gas effects. It was found that virtually unlimited shock wave acceleration coupled with precise locality of the front modification during the shock-interface interaction can be further amplified in some real gases. The level of the effect is on the order of 10-20% at the first moments of the interaction and it increases non-linearly with time.

To estimate the effect, first the *RH*-relations were reconsidered, in the range of gas parameters common for shocks and heated gas regions created in discharges. While a similar data already existed, the applications considered here required it be in a wider range of gas parameters and in a specific form, as a function of Mach number ratio. Non-dissociating gas in thermal equilibrium state on both sides of discontinuity with excited translational, rotational, vibrational, and electronic degrees of freedom was considered. For diatomic nitrogen at *T*=3000 *K,* an 11% density increase and 17% temperature and 7% pressure decrease, compared to the levels predicted by ideal-gas *R-H* relations, were found. The 5% correction to the speed of sound was determined and the correction level was strongly dependent on the stage and the way the wave is created. The correction to the jump relations was also found to be quite sensitive to the speed of sound correction. The size of the real-gas effect found here and the curve trends are rather typical for most real gases described in literature. As intermediate results, the specific heat correction $K_c$ and the corresponding specific heat ratio $\gamma_c$, for an extended range of gas temperatures in the nitrogen gas were obtained. The advantage of the procedure used here is that its relations are mainly based on experimental data thus avoiding the complexity accounting for all possible effects in real gas, yet offering a satisfactory precision level. All the experimental data used in the calculations, such as enthalpy, specific heat capacity, specific relaxation times, and characteristic temperatures for rotational, vibrational and electronic excitation degrees of freedom is conveniently available in literature. Since the data is usually given within definite ranges of temperature, the model applicability can be limited with this. However, if the desired data is outside of the range, the use of some analytical procedures presented here can be utilized as an alternative.

To apply the shock refraction equations to the specific gas conditions, real-gas *RH*-relations were used following the approach developed in [40]. Solving the corrected refraction equation together with the system of equations for the energy cumulation in the interaction yields the expressions for shock dynamics in the form of shock profiles and the speed. To sense the size of the effect, a numerical simulation was performed for the interaction between a plane shock wave and a plasma cloud of spherical or cylindrical geometry suspended in a colder surrounding gas, for a particular case of diatomic nitrogen at *T*=3000K. The most important result obtained here is that under real-gas conditions the extremely high energies acquired by a shock wave propagating down the decreasing density gradient can become even stronger. The reduced interface reflectivity and consequently increased shock transmission energy in real gas is the key phenomena explaining the effect.

Additional dimensions acquired by an initially flat front are due to the synergetic contribution of the interface curvature and the energy cumulation effect. These affect the front in both, transversal and (mostly) longitudinal directions, thus contributing to quickly increasing front distortions. As in ideal gas, the distortion quickly evolves into significant front stretching that is still localized along the symmetry axis and thus enabling precise locality in ignition applications. Due to an increased effect, the gas temperature behind it reaches very high values and at a higher rate.

With faster shock acceleration, the front portion around its tip arrives at the zero density plane sooner, while most of the remaining portions of the front continue its motion at much slower rates. When the whole front arrives at the final plane, the shock fully disappears followed by a jet-like formation of very hot gas exiting into a vacuum behind the plane. Note that at this point, when the particles and the radiation mean free paths becoming infinite, the gas-dynamical approximation and the model used here are no longer valid and the problem should be considered based on gas-kinetic methods. The boundary between the approximations can be set at the density of the order of $10^{-5}$ g/cm$^3$ when ordinary shocks can still be observable in experiments. Thus the supposedly infinite gas temperature behind the shock will be actually limited due to available dissipative process, usually radiation.

The results or approach described here can be also of interest in shock-flame interaction research, detonation, thermal energy deposition in the flow for hypersonic flight and space craft design, in astrophysics for understanding the dynamics of shock waves generated in stellar interior, or in fusion research where imploding shock waves in spherical and elliptical geometry are used.